\documentclass[proof]{WileyASNA-v1}

\articletype{Article Type}%

\received{26 April 2016}
\revised{6 June 2016}
\accepted{6 June 2016}

\begin{document}

\title{The crucial discovery of thermonuclear X-ray bursts: never throw away old data!}

\author[1]{Erik Kuulkers*}

\authormark{Erik Kuulkers}

\address[1]{\orgdiv{ESTEC}, \orgname{ESA}, \orgaddress{\state{Keplerlaan 1, 2201 AZ Noordwijk}, \country{The Netherlands}}}

\corres{*\email{Erik.Kuulkers@esa.int}}

\abstract{The detection of Type I X-ray bursts is attributed to those seen by the Astronomical Netherlands Satellite (ANS) in September 1975 from the globular cluster NGC\,6624 containing the X-ray source 4U\,1820$-$303. I revisit these X-ray bursts, by re-analysing data from the Soft X-ray Experiment (SXX) onboard ANS, which were stored on microfiche. Earlier accounts of X-ray bursts had been reported; the first Type I X-ray burst recorded is the one observed by Vela~5B from Cen\,X-4 in July 1969.}

\keywords{binaries: close, stars: neutron, X-rays: binaries, X-rays: bursts, history and philosophy of astronomy}

\maketitle

\section{Introduction}\label{sec1}

Canonical Type I X-ray bursts \citep{Hoffmanetal1978} result from thermonuclear shell flashes on a neutron star, which is caused by the ignition of either He and/or H-rich material supplied by a binary companion star (\citealp{HansenandvanHorn1975}; \citealp{WoosleyandTaam1976}; \citealp{MaraschiandCavaliere1977}; \citealp{LambandLamb1978})\footnote{For reviews on Type I X-ray bursts, see, e.g., (\citealp{Lewinetal1993}; \citealp{StrohmayerandBildsten2006}; \citealp{GallowayandKeek2021}.}. These bursts generally appear as short transient events wherein the X-ray intensity rises rapidly on a time scale of seconds, and decays in an exponential fashion back to the pre-burst level. The decay lasts almost always longer than the rise. Burst durations range from several seconds up to half an hour.
The energy spectra during these events can be fitted with a black-body model with temperatures of the order of $\sim$1\,keV
with an emission size of about 10\,km, consistent with a neutron star \citep{Swanketal1978}.
Generally, the burst energy spectra harden during the rise and peak, and soften during the decay.

\section{Astronomical Netherlands Satellite: ANS}\label{sec2}

\subsection{ANS instrumentation}

The first extra-Solar source discovered in X-rays was Scorpius X-1, during a rocket flight in 1962 \citep{Giacconietal1962}.
Starting with the first dedicated X-ray astronomy satellite, Uhuru \cite[e.g., ][]{Giacconietal1971}, many satellites in the 1970's with X-ray detectors were mostly scanning satellites. 
The Astronomical Netherlands Satellite (ANS), launched on 30 August 1974, was the first 3-axis stabilized X-ray satellite, and so could point to sources for longer stretches of time compared to previous satellites \citep{BloemendalandKramer1973}. ANS was active from 1974--1976.

ANS had three instruments onboard:
\begin{itemize}
\item Soft X-ray Experiment \cite[SXX; ][]{Brinkmanetal1974,denBoggendeandLafleur1975}. It consisted of a grazing-incidence paraboloidal-mirror with
a proportional counter in the focal plane covering the energy range around 0.25\,keV and a proportional counter
sensitive at 0.3--0.5\,keV and 1--7\,keV. The time resolution ranged from 0.125\,s (1 energy band) to 16\,s (7 energy channels)
\item Hard X-ray Experiment \cite[HXX; ][]{Gurskyetal1975}. It consisted of a large area detector (LAD) 
unit for measuring 1--30\,keV X-rays, and a Bragg-crystal spectrograph tuned for detection of the silicon lines.
\item UV-telescope sensitive at 150--330\,nm \citep{vanDuinenetal1975}.
\end{itemize}

\begin{figure*}[t]
\centerline{\includegraphics[width=492pt]{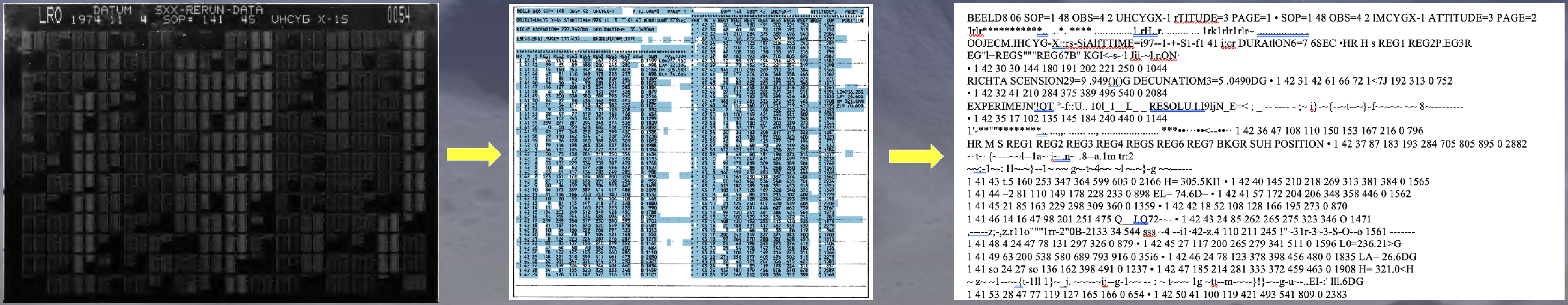}}
\caption{Example microfiche (left), which was scanned (middle) and OCR-ed (right).\label{microfiche}}
\end{figure*}

\begin{figure*}[t]
\centerline{\includegraphics[height=342pt,angle=-90]{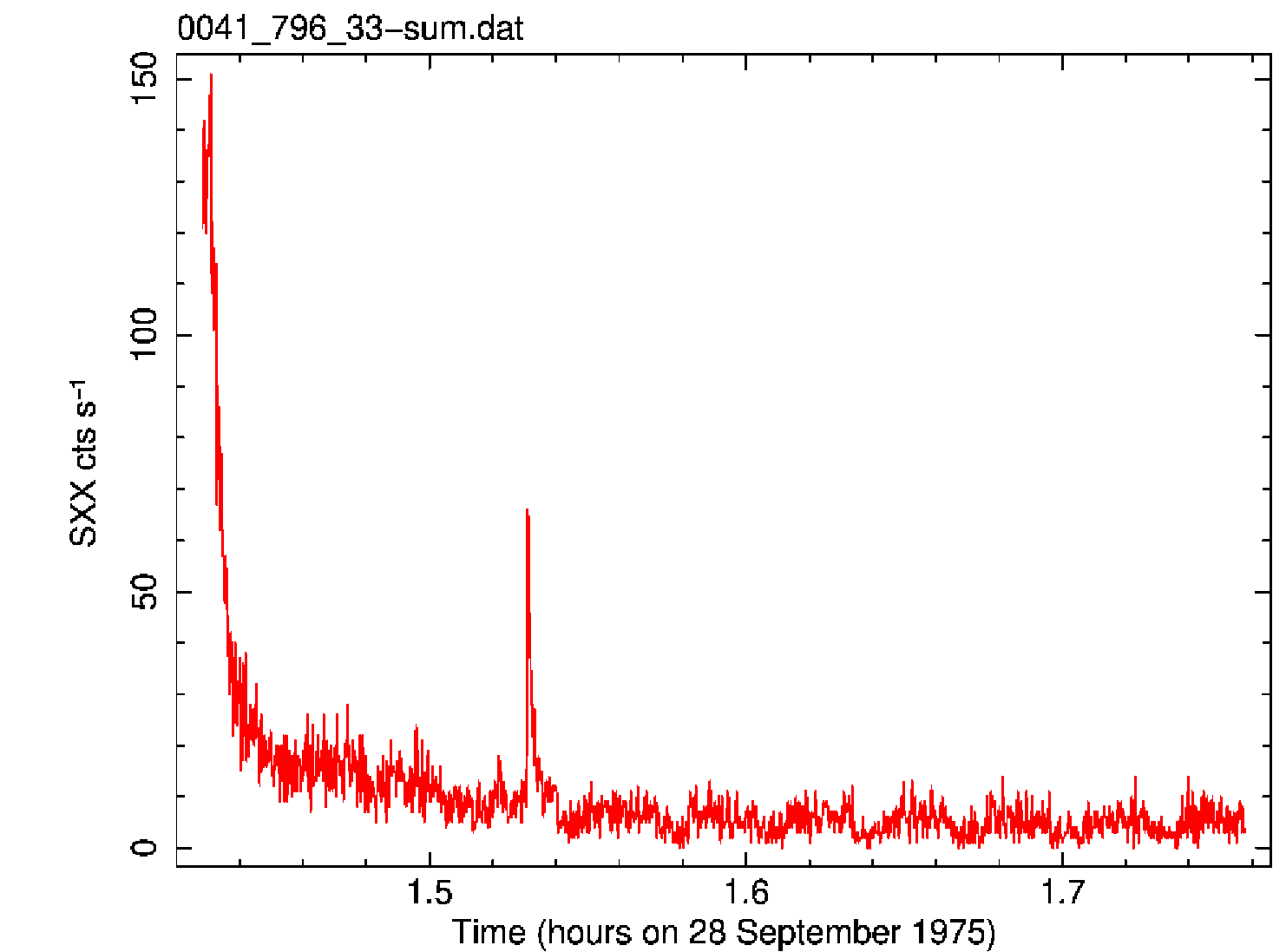}}
\caption{SXX light curve (1--7\,keV, uncorrected for background) during one of the passages on 28 September 1975 when ANS pointed towards NGC\,6624 (4U\,1820$-$303). Data from Satellite Operations Program (SOP) 796, Observation (OBS) 33, total 1186\,s exposure. The strong decay in flux at the start is due to the satellite exciting the radiation belts. The satellite was rocking between the source and background next to the source with an offset of 65' (see also Fig.~3, top panel); this results in a wavy structure of the light curve.\label{fig2}}
\end{figure*}

\begin{figure}[t]
\centerline{\includegraphics[width=78mm]{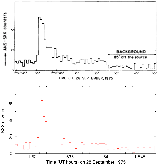}}
\centerline{\includegraphics[width=80mm]{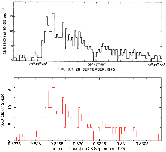}}
\caption{Comparison of SXX light curves (1--7\,keV) of the two X-ray bursts from 4U\,1820$-$303 on 28 September 1975, as reported by Grindlay et al.\ (1976; 1st and 3rd panel) and from re-analysis of the SXX data (2nd and 4th panel). The subtle differences are due to reprocessing of the data (see Sect.~2.1). 1st X-ray burst: SOP=796, OBS=33, rocking with offset 65', time resolution 1\,s, total exposure 1186\,s; 2nd X-ray burst: SOP=797, OBS=9, time resolution 0.125\,s, total exposure 257\,s.\label{fig3}}
\end{figure}

\subsection{ANS observations of 4U\,1820$-$303}

During a rocket flight in 1974 the black-hole X-ray binary Cygnus X-1 showed msec bursts \citep{Rothschildetal1974}.
At the time, \cite{BahcallandOstriker1975} suggested that the globular cluster NGC\,6624 contained a central black hole of mass $\sim$100--1000\,M$_{\odot}$ based on the soft X-ray spectrum and variability. 
It was postulated that if a Solar-mass black hole shows msec bursts, a super-massive black hole of $\sim$1000\,M$_{\odot}$ might show second-long bursts (see \citealp{Heise2010}). 
It was thus decided to spend time with ANS on observing NGC\,6624, which contained the X-ray source 4U\,1820$-$303, detected earlier with Uhuru \citep{Giacconietal1974}. 
4U\,1820$-$303 is nowadays known as a low-mass X-ray binary at a distance of about 8\,kpc with an an orbital period of only 11.4\,min. 
It contains a 0.06–-0.07\,M$_{\odot}$ helium white dwarf with a neutron star (see, e.g., \citealp{Yuetal2024}, and references therein).

Various ANS observations towards NGC\,6624/4U\,1820$-$30 were done in September 1974, and March and September 1975.
During observations on 28 September 1975 (see Figs.~2 and 3), two X-ray bursts were detected \citep{GrindlayandHeise1975,Grindlayetal1976}.
Each displayed a rapid rise in flux ($<$1\,s) by a factor of 20--30 followed by a $\sim$8\,s exponential decay.
A fit to the HXX burst data with a thermal spectrum revealed a black-body temperature of $kT = 5(+3,-2)$\,keV at the peak of the burst, while during the decay it became $kT = 20(\pm 8)$\,keV. 
Adopting a distance of 10\,kpc, the X-ray luminosity increased from $\sim$3$\times$10$^{37}$erg\,s$^{-1}$ to $\sim$1$\times$10$^{39}$\,erg\,s$^{-1}$. The total burst energy was $\sim$2$\times$10$^{39}$\,erg \citep{Grindlayetal1976}.

The decay portion and progressive spectral hardening of the ANS bursts from NGC\,6624 had been interpreted as a result of scattering of a brief primary burst of relatively 
soft X-rays by a surrounding hot-plasma cloud (\citealp{GrindlayandGursky1976b}; \citealp{Canizares1976}; \citealp{Grindlay1978}). The X-rays would originate from the vicinity of a supermassive 1000\,M$_{\odot}$ black hole, 
and thus confirming the supermassive black-hole hypothesis in NGC\,6624. Other explanations were also put forward (see, e.g., \citealp{LewinandJoss1981}, and references therein).

However, it was soon realized that these X-ray bursts are actually due to thermonuclear runaways on a neutron star (see Sect.~1).
The occurrence of X-ray bursts in 4U\,1820$-$303 depends on the state of the binary: X-ray bursts cease when its X-ray intensity transits from the so-called low to the hard state 
\citep{Clarketal1977}\footnote{X-ray states correspond to different accretion regimes, i.e., the accretion rate is thought to increase from low to high state.}.
Indeed, as judged from the Vela~5B (3--12\,keV) data \citep{PriedhorskyandTerrell1984}, the September 1975 ANS observations were taken during a low state.

The X-ray satellite SAS-3 was launched in May 1975. After the report of the events seen by ANS, bursts of X-rays, recurrent at nearly equal intervals, 
were detected in data obtained during an observation of 4U\,1820$-$303 on 17--19 May 1975. 
The properties were similar to that seen by ANS. The average X-ray intensity of 4U\,1820$-$303 during these observations was also low during the X-ray bursts \citep{Clarketal1976}.

If the bursts seen by ANS are indeed Type I X-ray bursts, the apparent increase of $kT$ seems to be a problem. 
The expected cooling during decay of a Type I X-ray burst is difficult to prove with the SXX data, because of lack of statistics. 
The X-ray bursts may need a reanalysis; the instrument response info is still available in paper form (e.g., \citealp{denBoggende1979}). 
Also, a comparison with those bursts seen later on with more sensitive instruments is warranted.

The SXX data has been preserved on 157 microfiches; each microfiche contains basically 208 pages of ASCII data. 
During November 1975 to February 1976, 1800 tapes with the ANS data of all earlier passages were re-processed, to profit from the improvements in the processing software; these were labeled "re-run" data \citep{Lamers2016}. 
At the end of 2023 all microfiches were scanned and OCR-ed; see Fig.~\ref{microfiche} for an example. The OCR-ed results of the relevant data were manually inspected and corrected.
To date, the data from the HXX are not anymore available.\footnote{Note that
the NASA Space Science Data Archive, NSSDC, states that the HXX data are “stored on one-half inch, digital magnetic tape produced on a NOVA 3 computer system” and that the “data set can be accessed by using the computer reduction facilities at HCO/SAO” (NSSDCA ID: ASXR-00038). However, the information is obsolete as these appear to be not available anymore (Dave Williams, Josh Grindlay, 2024, private communication).}

\section{Are those seen by ANS the first X-ray bursts?}

Bursts with timescales of seconds were discovered using data from two Vela-5 spy satellites designed to detect nuclear tests (Klebesadel et al.\ 1973); we now know that of most these are related to gamma-ray bursts (GRBs). 
However, \cite{Belianetal1972} had reported a burst in X-rays on 7 July 1969 with Vela~5B coming from the X-ray transient Cen\,X-4, with a duration of about 10\,min.
It showed a fast rise and a power-law decay. The hardness ratio (6--12\,keV over 3--6\,keV) softened during the decay. 
The burst emission could be modeled with a Planckian spectrum, with $kT \simeq 2.6$\,keV at the peak and $kT \simeq 1.3$\,keV during the decay, consistent with the observed softening. 
At the peak, the burst reached a flux which was 60 times brighter than that seen from Crab (!), corresponding to a luminosity of about 2$\times$10$^{38}$\,erg\,s$^{-1}$ (at the source distance of 1.2\,kpc; \citealp{Kuulkersetal2009}). 
The total energy release was about 5$\times$10$^{39}$\,erg (at 1.2\,kpc; Matsuoka et al.\ 1980). The X-ray burst was interpreted by Belian et al.\ (1972) as an accretion event.
However, although longer than the typical Type I X-ray burst, it resembles the so-called intermediate-duration Type I X-ray bursts (see, e.g., \citealp{Alizaietal2023}, and references therein).

A subset of the Vela-5 bursts were concentrated in the region of the Norma constellation \citep{Belianetal1976}; they probably originated from the X-ray bursters 4U\,1608$-$52 and/or from 4U\,1636$-$536 \citep{LewinandJoss1981}.
Re-analysis of Uhuru data between 1970 and 1973 revealed four burst events (in December 1971 and May 1972) which also originated from the same region in Norma (\citealp{GrindlayandGursky1976a,GrindlayandGursky1976c}). 
No enhanced emission was seen during Uhuru transits either before or after the events.
Two of the Uhuru events show a rise time of $<$2\,s, spectral hardening with time, and peak intensity comparable to
that of the Crab. For the other two events no evidence for a spectral change during the burst was observed.

It was soon found out that comparable short bursts like those seen by ANS had been reported from observations by the Russian Kosmos~428 satellite in June 1971 (\citealp{Babushkinaetal1975a}; \citealp{Babushkinaetal1975b}).
While these bursts were very similar to the ANS X-ray bursts in their time
profile and spectral hardening, their much higher (factor 10$-$100) energy flux and
detected energies originally suggested a closer relationship to GRBs \citep{Grindlay1976}.
Indeed, the X-ray instrument on-board Kosmos~428 was sensitive between 40--190\,keV. Since Type I X-ray bursts reach temperatures up to $kT \sim 3$\,keV, they will not show up typically above about 30\,keV (see, e.g., \citealp{Kuulkersetal2010}). 
They are nowadays not believed to be of cosmic origin, but may, e.g., be caused by the Earth's magnetosphere \cite{Lewin1977}\footnote{\cite{Heise2010} mentions that Lewin correlated it with the satellites geoposition in 1977, who then concluded that the bursts occurred near the Van Allen belts.}). 
Another postulation is that they are due to (other) Russian satellites that were powered by nuclear electrical generators, which contain extremely strong radioactive sources and would shower the detectors with gamma-rays emitted by the radioactive sources when they came near each other \citep{LewinandGoldstein2011}. Note that the orbital elements of the relevant satellites are still available; a re-analysis of their orbits could give more conclusive evidence of the orgin of the bursts.

An X-ray burst with duration of 15\,s was detected on 19 September 1972 by Uhuru, coming from the globular cluster NGC\,1851 \citep{JonesandForman1976}, which contains the known X-ray burster MX\,0513$-$40. 
The X-ray intensity of the source was relatively constant, although the spectrum was observed to harden significantly.

Two (possibly four) soft X-ray events were observed by the Apollo~15 (August 1971) and Apollo~16 (April 1972) X-ray spectrometers during the trans-Earth coast (\citealp{Gorensteinetal1974}). Their origin is not clear, but one event may be an X-ray burst (see \citealp{Kuulkersetal2009}, for a discussion).
Note that a GRB was observed by the gamma-ray and X-ray spectrometers on Apollo~16 as well as by Vela~6A on 27 April 1972 \citep{Metzgeretal1974,Trombkaetal1974}. 

Based on the literature, the event seen by Vela~5B in July 1969 \citep{Belianetal1972}, is the earliest record of a (Type I) X-ray burst.

\section{Conclusions}

The 1970’s data are (not only) of historical value; old X-ray data can still be 
a rich resource many decades after. They enable us to compare results with those of much later
measurements and interpret them in light of what we know nowadays. The data from the 1970’s extend the long-term baseline for characterizing state change and burst behaviour in sources like 4U\,1820$-$303. 

The first Type I X-ray burst ever observed is the one from Cen\,X-4, recorded on 7 July 1969 by the Vela~5B satellite.

Last but not least: never throw away old data; (legacy) archives are important!


\section*{Acknowledgments}

I want to thank Bert Brinkman for providing me the ANS/SXX data on microfiche when he retired. The microfiches were scanned to PDF by MEDIAFIX in October 2023.

\bibliography{ANS-XRBs}%

\end{document}